\documentclass[twocolumn]{aastex631}

\usepackage{graphicx}
\usepackage{dcolumn}
\usepackage{bm}
\usepackage{hyperref}
\usepackage{mathtools}
\usepackage{array}
\usepackage{tabularx}

\shorttitle{Constraints on coasting cosmologies from GW sirens}

\begin{document}

\title{Constraints on coasting cosmological models from gravitational-wave standard sirens}

\author[0000-0001-7576-0141]{Peter Raffai}
\affiliation{Institute of Physics and Astronomy, E\"otv\"os Lor\'and University, 1117 Budapest, Hungary}
\affiliation{HUN-REN–ELTE Extragalactic Astrophysics Research Group, 1117 Budapest, Hungary}

\author[0000-0001-5942-0470]{M\'aria P\'alfi}
\affiliation{Institute of Physics and Astronomy, E\"otv\"os Lor\'and University, 1117 Budapest, Hungary}

\author[0000-0003-3258-5763]{Gergely D\'alya}
\affiliation{Department of Physics and Astronomy, Universiteit Gent, B-9000 Ghent, Belgium}

\author[0000-0002-5556-9873]{Rachel Gray}
\affiliation{SUPA, University of Glasgow, Glasgow, G12 8QQ, United Kingdom}

\correspondingauthor{Peter Raffai}
\email{peter.raffai@ttk.elte.hu}

\begin{abstract}
We present the first test of coasting cosmological models with gravitational-wave standard sirens observed in the first three observing runs of the LIGO-Virgo-KAGRA detector network. We apply the statistical galaxy catalog method adapted to coasting cosmologies and infer constraints on the $H_0$ Hubble constant for the three fixed values of the curvature parameter $k=\left\{ -1,0,+1 \right\}$ in $H_0^2 c^{-2}$ units. The maximum posteriors and $68.3\%$ highest density intervals we obtained from a combined analysis of $46$ dark siren detections and a single bright siren detection are $H_0=\left\{68.1^{+8.5}_{-5.6},67.5^{+8.3}_{-5.2},67.1^{+6.6}_{-5.8} \right\}~\mathrm{km\ s^{-1}\ Mpc^{-1}}$, respectively. All our constraints on $H_0$ are consistent within one sigma with the $H_0$ measured with the differential age method, which provides a constraint on $H_0$ in coasting cosmologies independently from $k$. Our results constrain all cosmological models with $a(t)\propto t$ linear expansion in the luminosity distance and redshift range of the $47$ LIGO-Virgo detections, i.e. $d_\mathrm{L}\lesssim 5~\mathrm{Gpc}$ and $z\lesssim 0.8$, which practically include all (both strictly linear and quasi-linear) models in the coasting model family. As we have found, the coasting models and the $\Lambda$CDM model fit equally well to the applied set of gravitational-wave detections.
\end{abstract}

\section{Introduction} 
\label{sec:intro}

Coasting cosmologies is a family of cosmological models with the common feature that the $a(t)$ scale factor grows linearly with cosmic time $t$ (for a review, see~\citealt{Casado_2020}). Such models include ones suggesting strictly $a(t)\propto t$ linear expansion for the universe from the Big Bang to the present cosmic time, while in quasi-linear models the universe follows an evolution similar to the one in the current concordance model of cosmology (termed Lambda Cold Dark Matter or $\Lambda$CDM model, see~\citealt{Peebles_Ratra_2003} for a review) at early times and smoothly transitions to linear expansion around a late time and redshift $z_\mathrm{c}<z_{*}$, where $z_*$ is the redshift at recombination~\citep{Planck_2018}. Members of the coasting model family differ in the physical principles or mechanisms they propose as being responsible for the linear expansion, and/or in the value of the $k$ spatial curvature they suggest or allow. For example, the dynamics proposed by the earliest coasting model, developed by Arthur Milne in the 1930s~\citep{Milne_1935}, resembles that of an empty ($\rho=0$) universe with zero $\Lambda$ cosmological constant and negative $k$. A more recent example for a universe with linear expansion and $k=-1$ is given by the Dirac-Milne model~\citep{Benoit_Chardin_2012}. Other coasting models, such as the $R_{\mathrm{h}}=ct$ model~\citep{Melia_2007,Melia_Shevchuk_2012,Melia_2020a} and the eternal coasting model by John and Joseph~\citep{John_Joseph_1996,John_Joseph_2000} suggest $k=0$ and $k=+1$, respectively, although their core postulates allow any other value for $k$ (see e.g.~\citealt{John_Joseph_2000,John_Joseph_2023}). 

There are both theoretical and empirical motivations for studying coasting models even in view of the yet unparalleled success of the $\Lambda$CDM model. Coasting models provide natural solutions to several known theoretical problems in the $\Lambda$CDM model, including the horizon, the flatness, the cosmological constant, the synchronicity, the cosmic coincidence and the cosmic age problems (see~\citealt{Casado_2020} for a review). Note however, that the horizon and flatness problems are solved by the $\Lambda$CDM model when extended with the theory of cosmic inflation~\citep{Guth_1981,Baumann_2009}, while others in the list may simply be unlikely coincidences in the realizations of $\Lambda$CDM model parameters rather being problems in the model itself. Yet, the recently confirmed tensions between the $H_0$ Hubble constant~\citep{Riess_2020} and the $S_8$ structure growth parameter~\citep{Di_Valentino_et_al_2021} measured locally and determined from cosmic microwave background (CMB) observations using the $\Lambda$CDM model~\citep{Planck_2018}, as well as other anomalies~\citep{Perivolaropoulos_et_al_2021}, may also signify the need for studying alternatives to the current concordance model of cosmology. Coasting models fit remarkably well to a wide range of cosmological datasets at low ($z\ll z_*$) redshifts (see e.g. in Table 2 of~\citealt{Melia_2018} and references therein). Strictly linear models however have difficulties in explaining the observed abundances of light chemical elements presumably set by the process of primordial nucleosynthesis in the early universe~\citep{Kaplinghat_et_al_1999,Sethi_et_al_1999,Kaplinghat_et_al_2000,Lewis_et_al_2016}, as well as the origin and properties of anisotropies observed in the CMB (see e.g.~\citealt{Fujii_2020,Melia_2020b,Melia_2022}), both of which are well elaborated and understood in the framework of the $\Lambda$CDM model~\citep{Dodelson_2003}. Another limitation of coasting models is that the new physics they propose is testable only on cosmological scales, and thus far they lack predictions that are within the reach of laboratory scale experiments.

Since achieving the first detection of gravitational waves (GWs) in 2015~\citep{Abbott_et_al_2016}, the Advanced LIGO~\citep{Aasi_et_al_2015}, Advanced Virgo~\citep{Acernese_et_al_2015}, and KAGRA~\citep{Akutsu_et_al_2021} detectors have completed three observing runs, detecting a total of $90$ GW signals from coalescing compact binaries~\citep{Abbott_et_al_2021a}. The detections included GW170817~\citep{Abbott_et_al_2017a}, a GW signal from a binary neutron star merger for which electromagnetic counterparts in various bands have also been found~\citep{Abbott_et_al_2017b}. The exact localization of the optical counterpart allowed the identification of the host galaxy of this event~\citep{Abbott_et_al_2017b}, and the precise determination of its cosmological redshift~\citep{Abbott_et_al_2017c}, earning the term {\it bright siren} for the GW source. So far, GW170817 has remained the only GW signal with an associated host, with all others originating from {\it dark sirens}, i.e. coalescing compact binaries with detected GW emissions but no EM counterparts. As~\citet{Schutz_1986} pointed out, the $d_\mathrm{L}$ luminosity distances of coalescing compact binaries can be inferred from their GW signals without the need for a distance calibrator, which makes them what we call {\it standard sirens}~\citep{Holz_Hughes_2005}. Standard sirens with identified host galaxies or with a set of possible host galaxies can be used to test the $d_\mathrm{L}(z)$ redshift versus distance relationship of a selected cosmological model, as well as to constrain the model parameters, most prominently the rate of expansion at present time, i.e. the $H_0$ Hubble constant~\citep{Dalal_et_al_2006,MacLeod_et_al_2008,Nissanke_et_al_2013}. Such constraints on parameters of the $\Lambda$CDM model have already been published by the LIGO-Virgo-KAGRA Collaboration~\citep{Abbott_et_al_2017c,Soares_Santos_et_al_2019,Abbott_et_al_2021b,Abbott_et_al_2023a}.

In this paper, we present the first attempt to use GW standard sirens for testing coasting cosmologies, and to infer $H_0$ from GW signals assuming an $a(t)\propto t$ coasting evolution of the universe within the redshift range of GW detections. Note that for a fixed $k$ curvature, $H_0$ is the only parameter of coasting models determining the redshift-distance relation, whereas in the $\Lambda$CDM model we need at least one additional parameter (typically the $\Omega_\mathrm{m}$ present-day matter density parameter) to describe this relationship. As a consequence, GW standard sirens provide tighter constraints on $H_0$ in coasting models and a more direct way for testing these models compared to the case of the $\Lambda$CDM model.

Our paper is organized as follows. In Section~\ref{sec:Data_and_Analysis} we describe the analysis, and the GW and galaxy data we used for our test. In Section~\ref{sec:Results} we discuss the results of our analyses. Finally in Section~\ref{sec:Conclusions} we offer conclusions about our work and the possible ways of continuing it in the future.

Throughout this paper we use $\Omega_\mathrm{m}=0.3065$ and $k=0$ (and $H_0=67.9~\mathrm{km~s^{-1}~Mpc^{-1}}$ where needed) from~\citet{Ade_et_al_2016} for the $\Lambda$CDM model, to allow direct comparisons with results published in~\citet{Abbott_et_al_2023a}.

\section{Data and Analysis}
\label{sec:Data_and_Analysis}

For our tests, we used the publicly available GWTC-3 data~\citep{Abbott_et_al_2021a} and \texttt{gwcosmo} code (\citealt{Gray_et_al_2020}; for a more recent and enhanced version of the code, see~\citealt{Gray_et_al_2023}) to rerun the~\citet{Abbott_et_al_2023a} analysis using the statistical galaxy catalog method adapted to coasting cosmologies\footnote{\url{https://github.com/MariaPalfi/gwcosmo_coasting}}. This means that we applied the following relationship between the $d_\mathrm{L}$ luminosity distances of the GW sources and the $z$ cosmological redshifts of their host galaxies:
\begin{equation}\label{Eq:dL}
d_\mathrm{L}(z)=\frac{c}{H_0}\left( 1+z \right)
\begin{cases} 
\sinh(\ln(1+z)) &\textrm{for $k=-1$}\\
\ln(1+z) &\textrm{for $k=0$}\\
\left| \sin(\ln(1+z)) \right| &\textrm{for $k=+1$}\\
\end{cases}
\end{equation}
where we limited our tests to the three discrete cases of $k=\left\{ -1,0,+1 \right\}$ for the curvature parameter measured in $H_0^2 c^{-2}$ units (corresponding to $\Omega_0=\left\{ 0,1,2 \right\}$ density parameters today, respectively), $c$ being the speed of light in vacuum.

To allow direct comparisons with results published in~\citet{Abbott_et_al_2023a}, we analyzed the same $47$ GW events from the GWTC-3 catalog that were selected for testing the $\Lambda$CDM model there, with matched filter signal-to-noise ratio obtained by the LIGO-Virgo detector network $\mathrm{SNR}>11$ and Inverse False Alarm Rate $\mathrm{IFAR}>4~\mathrm{yr}$, taking their maximum across the different search pipelines. From this set of GW events, $46$ correspond to dark sirens, with GW170817 being the only one originating from a bright siren identified in galaxy NGC4993 at redshift $z=\left( 1.006\pm 0.055 \right)\times 10^{-2}$~\citep{Abbott_et_al_2017c}. 

Also similarly to~\citet{Abbott_et_al_2023a}, we used the \texttt{GLADE+}\footnote{\url{https://glade.elte.hu/}} full-sky catalog of over $22$ million galaxies and $750$ thousand quasars~\citep{Dalya_et_al_2022,Dalya_et_al_2018} to select potential host galaxies in our analysis for the dark siren events. Using the measured $K_s$-band luminosities of galaxies in \texttt{GLADE+} (where available), we applied the luminosity weighting described in~\citet{Abbott_et_al_2023a} in our analyses of dark sirens, i.e. we weighted each galaxy with a probability of being the host that is proportional to its $K_s$-band luminosity.

The code \texttt{gwcosmo} uses a Bayesian framework to infer the posterior probability on $H_0$ from the input GW events. The methodology of the code is explained in details in~\citet{Gray_et_al_2020}. We applied \texttt{gwcosmo} in the pixelated sky scheme~\citep{Gray_et_al_2022} with pixel size $0.2~\mathrm{deg}^2$ for analyzing the well-localized GW190814 event~\citep{Abbott_et_al_2020} and $3.35~\mathrm{deg}^2$ for all other events. We used the POWER LAW + PEAK source mass model~\citep{Talbot_Thrane_2018,Abbott_et_al_2023b} with the same population parameters used in~\citet{Abbott_et_al_2023a} to describe the primary black hole mass distribution. Also, we used the LIGO and Virgo detector sensitivities during the O1, O2, and O3 observing runs to evaluate GW selection effects. In all analyses, we inferred $H_0$ using a uniform prior in the interval $H_0\in [20,140]~\mathrm{km\ s^{-1}\ Mpc^{-1}}$. Note that these are the same run settings for \texttt{gwcosmo} that were used to produce results in the framework of the $\Lambda$CDM model in~\citet{Abbott_et_al_2023a}, but limited only to the standard case of the~\citet{Abbott_et_al_2023a} analysis using the most plausible settings. Thus we refer the reader to~\citet{Abbott_et_al_2023a} for a more detailed discussion about the rationale behind the run settings.

$H_0$ for coasting cosmological models can be determined in a curvature-independent way using the so-called \emph{cosmic chronometer} or \emph{differential age} (DA) method originally introduced in~\citet{Jimenez_Loeb_2002} and~\citet{Simon_Verde_Jimenez_2005}. This method takes advantage of the fact that $H(z)=-\dot{z}\left( 1+z \right)^{-1}$ for all cosmologies (including both the $\Lambda$CDM and coasting models) that satisfy $a(z)=(1+z)^{-1}$, and that $\dot{z}$ can in practice be approximated as $\dot{z}\approx \Delta z \Delta t^{-1}$, where $\Delta z$ and $\Delta t$ are the redshift and age differences of e.g. pairs of galaxies at around various $z$ redshifts. Passively evolving galaxies allow measuring their $\Delta t$ age differences from observed differences in their stellar populations, from which $H(z)$ can be determined with uncertainties typically dominated by uncertainties of the $\Delta t$ differential age measurement. \citet{Melia_Maier_2013} used this DA method to determine $H_0$ by fitting $H(z)=H_0(1+z)$ in coasting models to $19$ $H(z)$ measurements from~\citet{Simon_et_al_2005,Stern_et_al_2010,Moresco_et_al_2012}, and obtained $H_0=63.2\pm 1.6~\mathrm{km~s^{-1}~Mpc^{-1}}$ regardless of $k$. We updated their result by fitting the $H(z)$ formula in coasting models to the latest set of $32$ $H(z)$ measurements~\citep{Simon_et_al_2005,Stern_et_al_2010,Moresco_et_al_2012,Zhang_et_al_2014,Moresco_2015,Moresco_et_al_2016,Ratsimbazafy_et_al_2017,Borghi_et_al_2022} summarized in Table 1 of~\citet{Moresco_et_al_2022} using the public \texttt{emcee}\footnote{\url{https://gitlab.com/mmoresco/CCcovariance}}~\citep{Foreman_Mackey_et_al_2013} Markov Chain Monte Carlo code with the full statistical and systematic covariance matrix of the data. We obtained $H_0=62.41^{+2.95}_{-2.96}~\mathrm{km~s^{-1}~Mpc^{-1}}$, and used this $H_0$ as a reference for consistency checks of the $H_0$ posteriors we obtained from GW standard sirens for the coasting models. Note that in contrast to~\citet{Abbott_et_al_2023a}, we cannot use $H_0$ values obtained by the Planck and SH0ES teams for comparisons (see~\citealt{Planck_2018} and~\citealt{Riess_et_al_2022}, respectively), as both results rely on assumptions valid for the $\Lambda$CDM model but not for coasting models.

\section{Results}
\label{sec:Results}

The most distant GW events we analyzed have $d_\mathrm{L}\simeq \nobreak 5~\mathrm{Gpc}$~\citep{Abbott_et_al_2023a}, corresponding to $z=0.78$ in the $\Lambda$CDM model, and $z=\nobreak [0.76;~0.79;~0.83]$ for the $k=\left\{ -1,0,+1 \right\}$ coasting models with $H_0=62.41~\mathrm{km~s^{-1}~Mpc^{-1}}$, respectively. Thus we conclude that our analysis tests all cosmological models with an $a(t)\propto t$ coasting evolution in the redshift range $z\lesssim 0.8$. Since even quasi-linear models propose a coasting evolution in this redshift range, this means that our analyses based on GW standard sirens can test all models in the coasting model family. Note also, that the tested $z\lesssim 0.8$ redshift range includes $z\simeq 0.64$, when the universe switched from decelerating to accelerating expansion, and $z\simeq 0.30$, when the universe switched from matter-dominated to $\Lambda$-dominated expansion in the $\Lambda$CDM model, making GW standard sirens excellent tools for making comparisons between expansion histories proposed by the $\Lambda$CDM and the coasting models.

\begin{figure}
 	\includegraphics[width=\columnwidth]{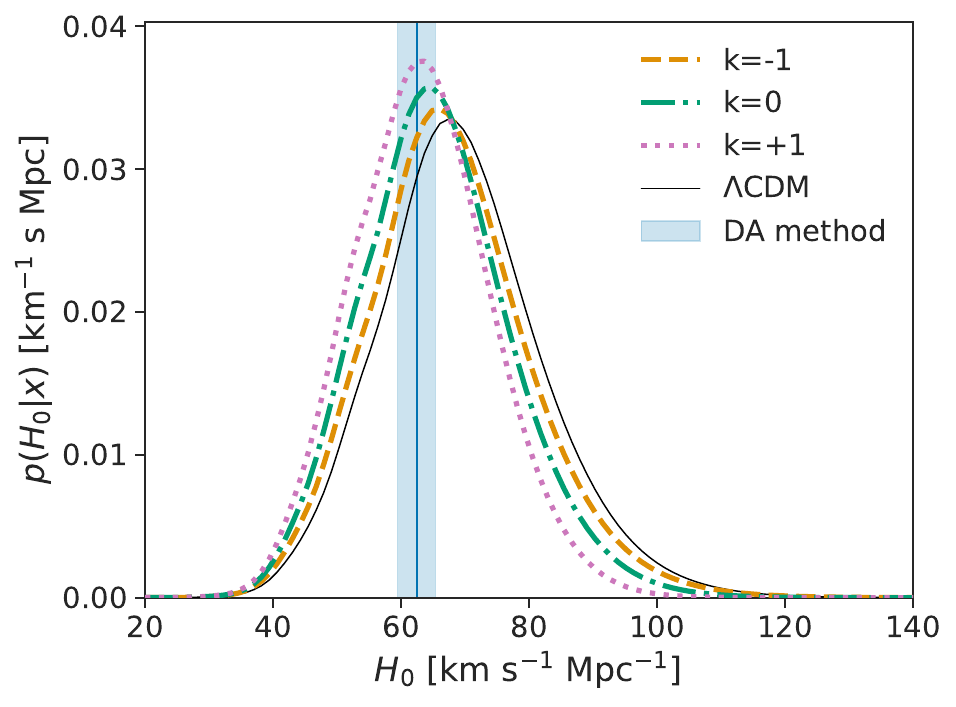}
     \caption{The GW measurements of $H_0$ from dark siren detections in the first three observing runs of the LIGO-Virgo-KAGRA detector network assuming coasting cosmologies with $k=\left\{ -1,0,+1 \right\}$ in $H_0^2 c^{-2}$ units, and the $\Lambda$CDM model. The maximum posteriors and $68.3\%$ highest density intervals for $H_0$ are given in Table~\ref{tab:table1}. We produced all posteriors using uniform priors in the interval $H_0\in [20,140]~\mathrm{km\ s^{-1}\ Mpc^{-1}}$. We also show our estimate of $H_0$ for coasting cosmologies using the differential age (DA) method, which is $H_0=62.41^{+2.95}_{-2.96}~\mathrm{km~s^{-1}~Mpc^{-1}}$ regardless of $k$.}
     \label{fig:fig1}
\end{figure}

\begin{figure}
 	\includegraphics[width=\columnwidth]{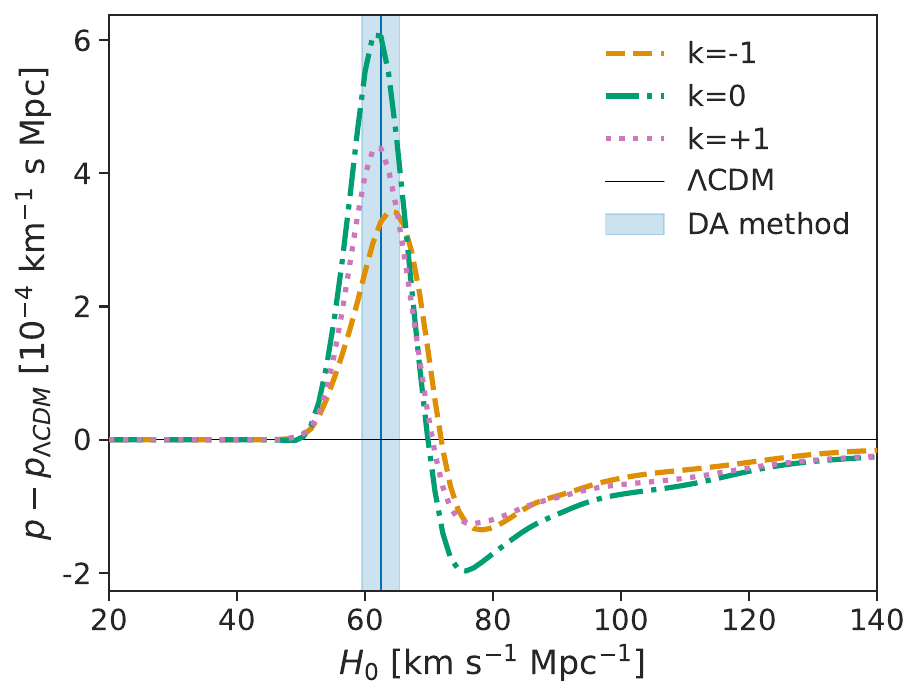}
     \caption{The GW measurements of $H_0$ from GW170817 (the only bright siren detection in the first three observing runs of the LIGO-Virgo-KAGRA detector network) shown in terms of differences between the $p$ posteriors for $H_0$ in various cosmologies (including $\Lambda$CDM, represented by the solid black line) and the $p_\mathrm{\Lambda CDM}$ posterior for $H_0$ in the $\Lambda$CDM model. The curves denoted by $k=-1$, $k=0$ and $k=+1$ correspond to coasting cosmologies with $k=\left\{ -1,0,+1 \right\}$ in $H_0^2 c^{-2}$ units. We give the maximum posteriors and $68.3\%$ highest density intervals for $H_0$ in Table~\ref{tab:table1}. We produced all posteriors using uniform priors in the interval $H_0\in [20,140]~\mathrm{km\ s^{-1}\ Mpc^{-1}}$. We also show our estimate of $H_0$ for coasting cosmologies using the differential age (DA) method, which is $H_0=62.41^{+2.95}_{-2.96}~\mathrm{km~s^{-1}~Mpc^{-1}}$ regardless of $k$.}
     \label{fig:fig2}
\end{figure}

\begin{figure}
 	\includegraphics[width=\columnwidth]{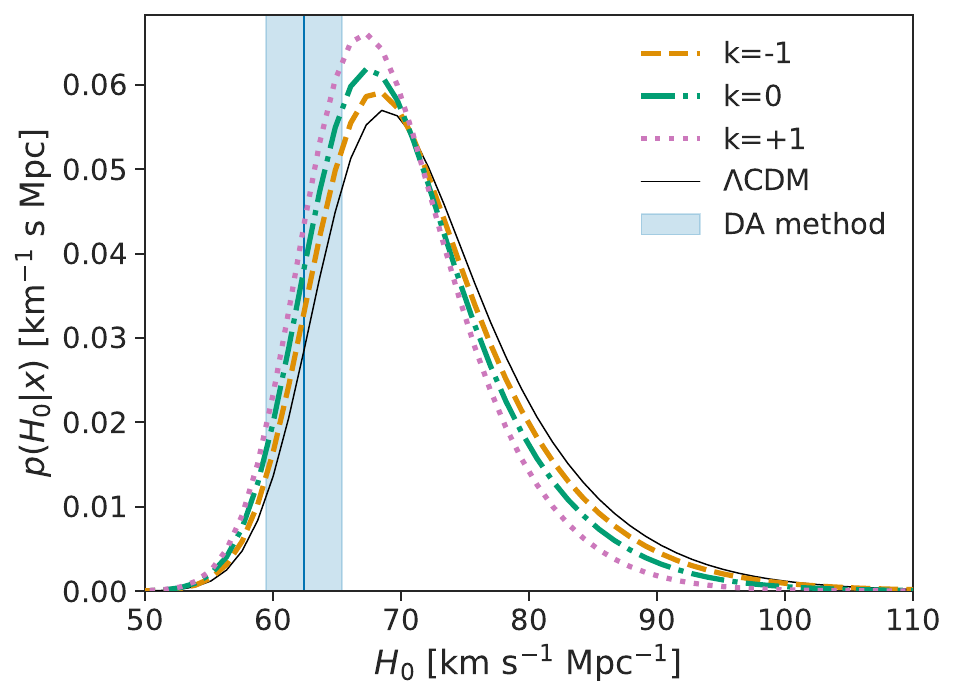}
     \caption{Combined posteriors for $H_0$ from the dark siren detections and the single bright siren detection (GW170817) in the first three observing runs of the LIGO-Virgo-KAGRA detector network. The curves denoted by $k=-1$, $k=0$, $k=+1$, and $\Lambda$CDM correspond to coasting cosmologies with $k=\left\{ -1,0,+1 \right\}$ in $H_0^2 c^{-2}$ units, and the $\Lambda$CDM model, respectively. The maximum posteriors and $68.3\%$ highest density intervals for $H_0$ are given in Table~\ref{tab:table1}. We produced all posteriors using uniform priors in the interval $H_0\in [20,140]~\mathrm{km\ s^{-1}\ Mpc^{-1}}$. We also show our estimate of $H_0$ for coasting cosmologies using the differential age (DA) method, which is $H_0=62.41^{+2.95}_{-2.96}~\mathrm{km~s^{-1}~Mpc^{-1}}$ regardless of $k$.}
     \label{fig:fig3}
\end{figure}

In Figure~\ref{fig:fig1}, we show the posterior distributions for $H_0$ in the $k=\left\{ -1,0,+1 \right\}$ coasting models and in the $\Lambda$CDM model we obtained from dark siren detections only. Due to the minor differences between the posterior distributions obtained from the single bright siren detection (GW170817), in Figure~\ref{fig:fig2} we show the differences between the posteriors for $H_0$ in various cosmologies (including $\Lambda$CDM) and the posterior for $H_0$ in the $\Lambda$CDM model. Finally, we show the combined posteriors for $H_0$ obtained from both dark and bright siren detections in Figure~\ref{fig:fig3}. We give the maximum posteriors and $68.3\%$ highest density intervals for $H_0$ in Table~\ref{tab:table1}, along with the logarithm of the Bayes factors between the tested cosmological models and the $\Lambda$CDM model, calculated for all detections. Note that these log Bayes factors are in the order of $10^{-7}-10^{-9}$ for the three coasting models, which in practice means that all four tested cosmological models fit equally well to the applied set of GW standard siren detections.

\begin{table}
\centering
\begin{tabularx}{1\columnwidth}{p{0.15\columnwidth}>{\centering}p{0.16\columnwidth}>{\centering}p{0.16\columnwidth}>{\centering}p{0.16\columnwidth}>{\centering\arraybackslash}p{0.15\columnwidth}}
\hline
\textbf{Model} & \textbf{Dark sirens} & \textbf{Bright siren} & \textbf{All sirens} & \boldmath$\log_{10}\cal{B}$ [$10^{-8}$] \\
\hline \hline
k=-1 & $65.9^{+12.9}_{-11.7}$ & $69.3^{+21.2}_{-8.1}$ & $68.1^{+8.5}_{-5.6}$ & -0.7\\
k=0 & $64.5^{+11.6}_{-11.7}$ & $69.3^{+21.3}_{-8.0}$ & $67.5^{+8.3}_{-5.2}$ & 11.0\\
k=+1 & $63.1^{+10.2}_{-11.5}$ & $69.3^{+21.2}_{-8.1}$ & $67.1^{+6.6}_{-5.8}$ & 12.4\\
$\Lambda$CDM & $67.7^{+13.0}_{-12.1}$ & $69.4^{+21.2}_{-8.2}$ & $68.7^{+8.4}_{-6.3}$ & 0\\
\hline
\end{tabularx}
\caption{\label{tab:table1}
The GW measurements of $H_0$ (maximum posteriors and $68.3\%$ highest density intervals in $\mathrm{km\ s^{-1}\ Mpc^{-1}}$ units) for coasting cosmologies with $k=\left\{ -1,0,+1 \right\}$ in $H_0^2 c^{-2}$ units, and for the $\Lambda$CDM model. The second and third columns indicate the $H_0$ measurements from dark siren detections and from the single bright siren detection (GW170817) in the first three observing runs of the LIGO-Virgo-KAGRA detector network. The fourth column shows the $H_0$ measurement from all these detections combined. We produced all posteriors using uniform priors in the interval $H_0\in [20,140]~\mathrm{km\ s^{-1}\ Mpc^{-1}}$. In the last column, we show the logarithm of the Bayes factors (in $10^{-8}$ units) between the tested cosmological models and the $\Lambda$CDM model, calculated for all GW siren detections.
}
\end{table}

The reason for the minor differences in $H_0$ posteriors for GW170817 is that, originating from a bright siren, both $d_\mathrm{L}=40^{+7}_{-15}~\mathrm{Mpc}$~\citep{Abbott_et_al_2023a} and $z=\left( 1.006\pm 0.055 \right)\times 10^{-2}$~\citep{Abbott_et_al_2017c} for the source are known. We can express $H_0$ in the coasting models ($H_{0,\mathrm{c}}$) and in the $\Lambda$CDM model ($H_{0,\Lambda}$) in terms of $d_\mathrm{L}$ and $z$ as
\begin{equation}\label{Eq:H0c}
H_{0,\mathrm{c}}\simeq \frac{c}{d_\mathrm{L}}z\left( 1+z \right)\left[ 1-\frac{1}{2}z+\mathrm{\cal{O}}(z^2) \right]
\end{equation}
\begin{eqnarray}\label{Eq:H0L}
H_{0,\Lambda} = &&\frac{c}{d_\mathrm{L}}\left( 1+z \right) \int_{0}^{z} \frac{\mathrm{d}z'}{\sqrt{\Omega_\mathrm{m}\left( 1+z' \right)^3+1-\Omega_\mathrm{m}}}\simeq \nonumber \\
&& \simeq \frac{c}{d_\mathrm{L}}z\left( 1+z \right)\left[ 1-\frac{3\Omega_\mathrm{m}}{4}z+\mathrm{\cal{O}}(z^2) \right],
\end{eqnarray}
and thus
\begin{equation}\label{Eq:H0s}
H_{0,\mathrm{c}}\simeq \left[ \frac{1-\frac{1}{2}z}{1-\frac{3\Omega_\mathrm{m}}{4}z}+\mathrm{\cal{O}}(z^2) \right]H_{0,\Lambda}
\end{equation}
which for $z\simeq 0.01$ and $H_{0,\Lambda}\simeq 69.4~\mathrm{km~s^{-1}~Mpc^{-1}}$ (see Table~\ref{tab:table1}) is $H_{0,\mathrm{c}}\simeq 69.2~\mathrm{km~s^{-1}~Mpc^{-1}}$, comparable to the maximum posteriors for $H_0$ in the coasting models in Table~\ref{tab:table1}.

Similarly to the results presented in~\citet{Abbott_et_al_2023a}, the $H_0$ values we obtained for dark sirens are dominated by the black hole population assumptions we used (see Section~\ref{sec:Data_and_Analysis}). As shown in~\citet{Abbott_et_al_2023a}, the main source of systematic uncertainty in this case is the choice of the peak location in the POWER LAW~+~PEAK mass model for primary black holes. For all values of $k$, our $H_0$ maximum posteriors would decrease with the peak shifting towards lower mass values, and vice versa. This systematic uncertainty can be reduced in the future by using a galaxy catalog in the analysis with a higher level of completeness, by constraining the black hole population model better with the increasing number of GW detections, or by jointly estimating parameters of the black hole population model alongside cosmological parameters~\citep{Mastrogiovanni_et_al_2023,Gray_et_al_2023}.

\section{Conclusions}
\label{sec:Conclusions}

In this paper, we have presented the first tests of coasting cosmological models with GW standard sirens. We applied the statistical galaxy catalog method with a version of the \texttt{gwcosmo} code we adapted to coasting cosmologies, and inferred constraints on $H_0$, the only parameter of coasting models with fixed values of $k=\left\{ -1,0,+1 \right\}$ in $H_0^2 c^{-2}$ units. We have presented the $H_0$ posteriors we obtained using $46$ dark siren detections in the first three observing runs of the LIGO-Virgo-KAGRA detector network (see Figure~\ref{fig:fig1}), using the single bright siren detection (GW170817, see Figure~\ref{fig:fig2}), and for all GW standard siren detections combined (see Figure~\ref{fig:fig3}). We have given the maximum posteriors and $68.3\%$ highest density intervals for $H_0$ in the selected cosmologies in Table~\ref{tab:table1}, along with the log Bayes factors between the tested models and the $\Lambda$CDM model, calculated for all GW siren detections. To check the consistency of our results with an independent measurement of $H_0$, we used $H_0=62.41^{+2.95}_{-2.96}~\mathrm{km~s^{-1}~Mpc^{-1}}$ as a reference, which we determined for coasting cosmologies independently from $k$ using the DA method. Our results test and constrain all cosmological models with $a(t)\propto t$ linear expansion in the luminosity distance and redshift range of the $47$ LIGO-Virgo detections, i.e. $d_\mathrm{L}\lesssim 5~\mathrm{Gpc}$ and $z\lesssim 0.8$, which practically include all (both strictly linear and quasi-linear) models in the coasting model family.

The log Bayes factors in Table~\ref{tab:table1} indicate that the coasting models and the $\Lambda$CDM model fit equally well to the applied set of GW standard siren detections. With the constraints on $H_0$ we derived, we have found that all $k=\left\{ -1,0,+1 \right\}$ coasting models are consistent within one sigma with the DA method value of $H_0$, and that there is an overall trend of the $H_0$ maximum posterior decreasing with increasing $k$ (thus, the maximum posterior for $k=+1$ is the closest to the $H_0$ measured with the DA method). Our measurements of $H_0$ however, with the large error bars, cannot set tight enough constraints on $k$ to exclude any of the three spatial geometries for coasting models from future considerations.

The growing number of GW standard siren detections with the current LIGO-Virgo-KAGRA network~\citep{Abbott_et_al_2018} soon to be expanded with LIGO-India~\citep{Saleem_et_al_2022}, as well as with future ground-based detectors such as Einstein Telescope~\citep{Punturo_et_al_2010} and Cosmic Explorer~\citep{Reitze_et_al_2019} will allow putting tighter constraints on $H_0$, with the potential of ruling out certain models in the coasting model family based on their inconsistency with the $d_\mathrm{L}(z)$ relation mapped out by GW standard sirens, or with independent determinations of $H_0$ and $k$. Additionally, alternative methods developed for measuring $H_0$ in the $\Lambda$CDM model with GW standard sirens (see e.g.~\citealt{Mastrogiovanni_et_al_2021,Mastrogiovanni_et_al_2023}; also, for a list of existing methods, see the conclusion section of ~\citealt{Abbott_et_al_2023a} and references therein) can be adapted in the future to testing coasting cosmologies and complementing results obtained with the \texttt{gwcosmo} implementation of the statistical galaxy catalog method.

\vspace{5mm}

The authors thank Juan Casado, Moncy V. John, Fulvio Melia, and Adam Riess for useful comments that improved the manuscript. This paper was reviewed by the LIGO Scientific Collaboration under LIGO Document P2300312. The authors are grateful for the support of the Cosmology and the Tests of $\Lambda$CDM subgroups of the LIGO-Virgo-KAGRA Collaboration. This material is based upon work supported by NSF's LIGO Laboratory which is a major facility fully funded by the National Science Foundation. The authors are grateful for computational resources provided by the LIGO Laboratory and supported by National Science Foundation Grants PHY-0757058 and PHY-0823459. This project has received funding from the HUN-REN Hungarian Research Network. This work makes use of \texttt{gwcosmo} which is available at \url{https://git.ligo.org/lscsoft/gwcosmo}.

\vspace{5mm}

\bibliography{GW_Coasting}{}
\bibliographystyle{aasjournal}

\end{document}